\documentclass[a4paper,11pt]{article}
\usepackage{pos}

\title{Measurement of the top quark pole mass using \ttbarjet\ events in the dilepton final state at $\sqrt{s}=$ 13 TeV}
\ShortTitle{Measurement of the top quark pole mass using \ttbarjet\ events in CMS}
\author*[1]{Sebastian Wuchterl}
\affiliation{Deutsches Elektronen-Synchrotron (DESY),\\
  Notkestrasse 85, Hamburg, Germany}
\note{For the CMS Collaboration.}
\emailAdd{sebastian.wuchterl@cern.ch}

\newcommand{\ttbar}{\ensuremath{\mathrm{t\bar{t}}}}
\newcommand{\ttbarjet}{\ensuremath{\mathrm{t\bar{t}}\text{+jet}}}
\newcommand{\ttbarnojet}{\ensuremath{\mathrm{t\bar{t}}\text{+0jet}}}
\newcommand{\fbinv}{\ensuremath{\mathrm{fb^{-1}}}}

\newcommand{\mt}{\ensuremath{\mathrm{m_{t}}}}
\newcommand{\mtpole}{\ensuremath{\mathrm{m_{t}^{pole}}}}
\newcommand{\mtmc}{\ensuremath{m_\mathrm{{t}}^\mathrm{{MC}}}}
\newcommand{\pt}{\ensuremath{p_\mathrm{{T}}}}

\abstract{
In these proceedings, a measurement of the differential cross section of top quark-antiquark pair (\ttbar) production in association with one additional jet (\ttbarjet) as a function of the inverse of the invariant mass of the \ttbarjet\ system, $\rho=340\,\text{GeV}/m_{\ttbarjet}$, is presented. The normalized \ttbarjet\ cross section is used to extract values for the top quark pole mass \mtpole\ by comparison to theoretical predictions at next-to-leading order accuracy.
The used data set corresponds to an integrated luminosity of 36.3\,\fbinv\ of proton--proton collisions as collected by the CMS experiment at $\sqrt{s}=$13 TeV. Events with two opposite-sign leptons are analyzed. Machine learning techniques are employed to improve the kinematic reconstruction of the main observable and the event classification. The unfolding to the parton level is performed using a profiled likelihood fit.
Given the ABMP16 parton distribution functions as a reference set, a value of $\mtpole=172.94\pm1.37\,\text{GeV}$ is extracted using the normalized \ttbarjet\ cross section.
}

\FullConference{%
  41st International Conference on High Energy physics - ICHEP2022\\
  6-13 July, 2022\\
  Bologna, Italy
}

\begin{document}
\renewcommand{\logo}{\relax}
\maketitle

\section{Introduction}
The top quark is the heaviest particle of the standard model (SM) of particle physics, and its mass \mt\ is an important input to global electroweak fits and calculations of the Higgs boson self-coupling. Its value has to be determined experimentally.
Direct measurements of \mt, which rely on the usage of multipurpose Monte-Carlo (MC) event generators, reach a precision on the order of 0.5 GeV. As they depend on the modeling of nonperturbative effects via the usage of heuristic models tuned to data, their interpretation is unclear. Usually, values for \mtmc\ are associated to the top quark pole mass, \mtpole, with an interpretation uncertainty on the order of 0.5--1 GeV~\cite{bib:hoangmass}.

In this contribution, a measurement of \mtpole\ by the CMS Collaboration~\cite{CMS-PAS-TOP-21-008} is presented, which was updated in Ref.~\cite{CMS:2022emx}. The production cross section of a top quark-antiquark pair (\ttbar) in association with at least one additional jet (\ttbarjet) as a function of the inverse of the invariant mass of the \ttbarjet\ system, $\rho=340\,\text{GeV}/m_{\ttbarjet}$, is sensitive to \mtpole.
It is measured at the parton level, and \mtpole\ is extracted from a comparison to theoretical predictions at next-to-leading-order (NLO) accuracy~\cite{bib:ttjPheno}. Large mass sensitivity is expected for values of $\rho>0.65$. The additional jet at parton level needs to have a transverse momentum of $\pt>30\,\text{GeV}$ and an absolute pseudorapidity of $|\eta|<2.4$.

\section{The experimental analysis}
Proton-proton collision events recorded by the CMS experiment, corresponding to an integrated luminosity of 36.3\,\fbinv, are analyzed. A combination of single and dilepton triggers is used to collect the events. They are selected if they contain one opposite-sign lepton pair with $\pt>25\,(20)\,\text{GeV}$ for the leading (subleading) lepton. Combinations of $\text{e}^{+}\text{e}^{-}$, $\mu^{+}\mu^{-}$, $\text{e}^{\pm}\mu^{\mp}$ are considered. Jets are selected with $\pt>30\,\text{GeV}$, and leptons and jets need to satisfy $|\eta|<2.4$. A tagging algorithm is used to identify jets originating from hadronized b quarks (b jets).
\begin{figure}[htbp]
\centering
\includegraphics[width=0.42\textwidth]{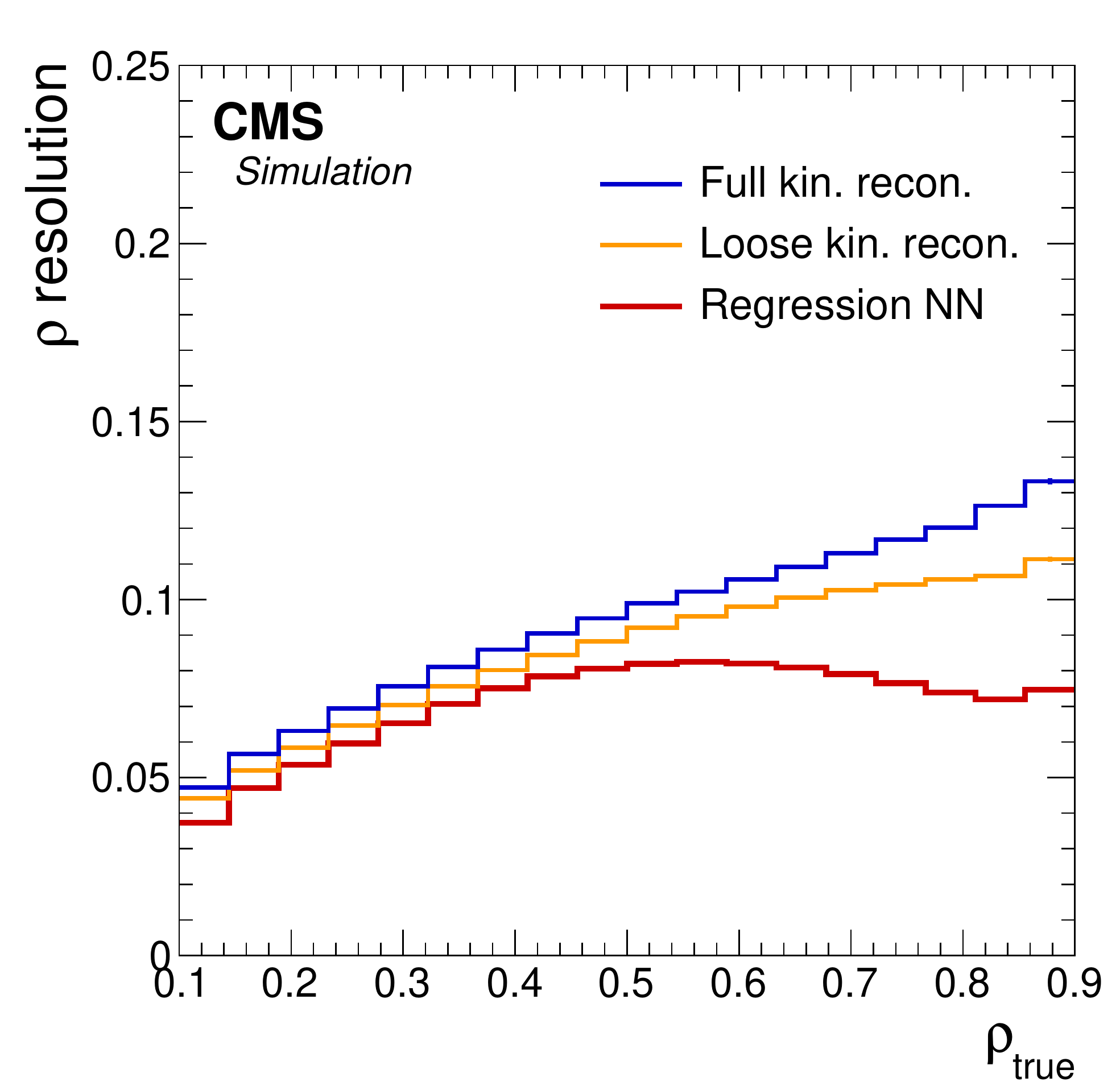}
\includegraphics[width=0.42\textwidth]{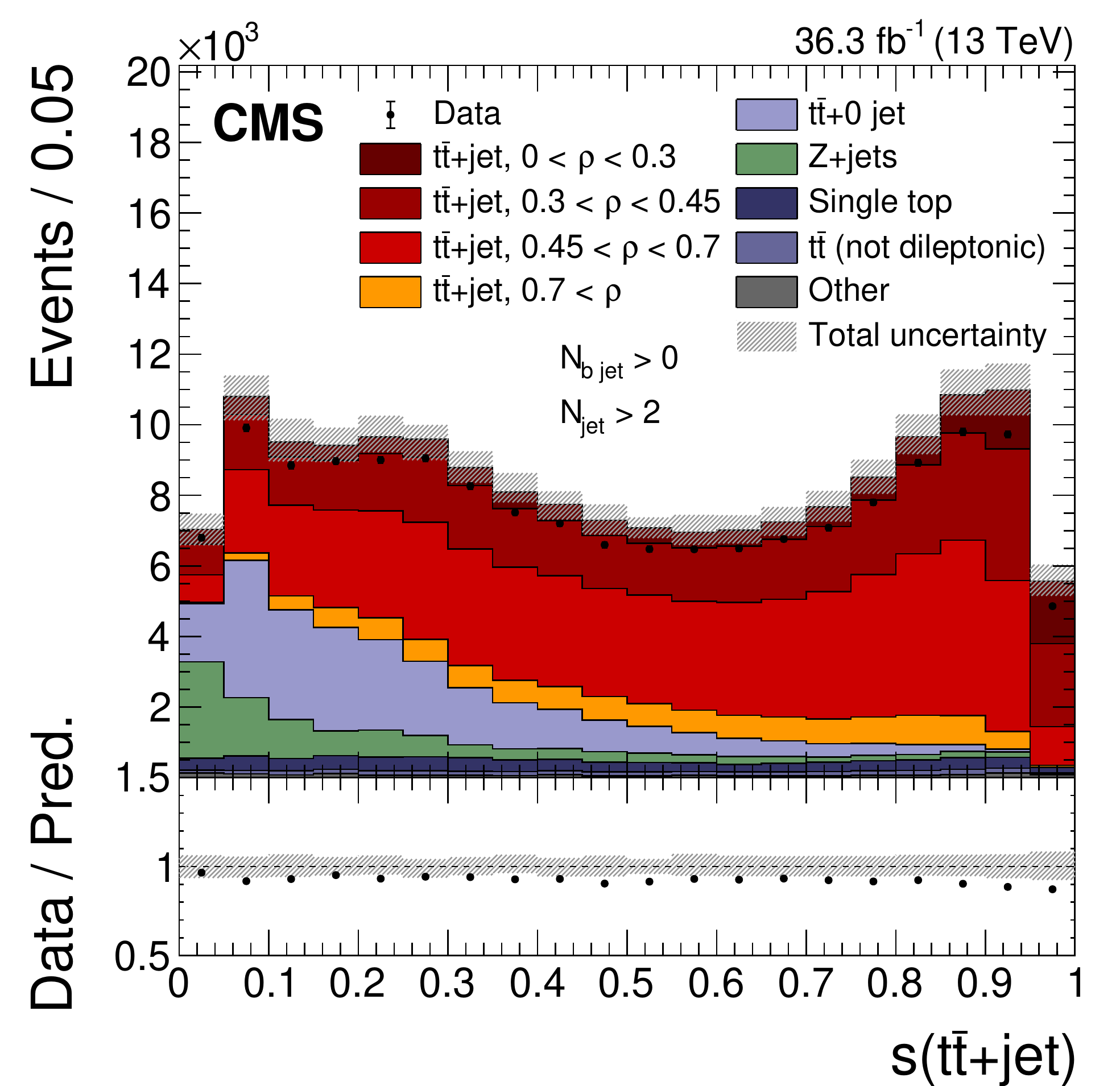}
\caption{The $\rho$ resolution as a function of the truth value for the regression NN and two analytical methods (left)~\cite{CMS:2022emx}. Data and MC predicted distributions as a function of the signal output node of the classifier NN~\cite{CMS:2022emx}.}
\label{fig:NN}
\end{figure}

In dileptonic \ttbarjet\ events, the amount of missing transverse momentum complicates the reconstruction of $\rho$. Thus, a multivariate technique is employed, and a regression neural network (NN) is trained. The resolution is shown in Fig.~\ref{fig:NN} left, and it is compared to approaches used in previous CMS publications. Using the NN approach, the $\rho$ resolution is improved by up to a factor of two depending on its value.

The dominant background contribution arises from \ttbar\ production without an additional jet, \ttbarnojet, where the reconstructed jet originates from pileup effects or phase space migrations. As this is particularly relevant for large values of $\rho$, a multiclass classification NN is designed to discriminate between signal and background processes. The signal response of the NN is shown in Fig.~\ref{fig:NN} right, where the data is compared to the MC expectations.

\begin{figure}[htbp]
\centering
\includegraphics[width=0.85\textwidth]{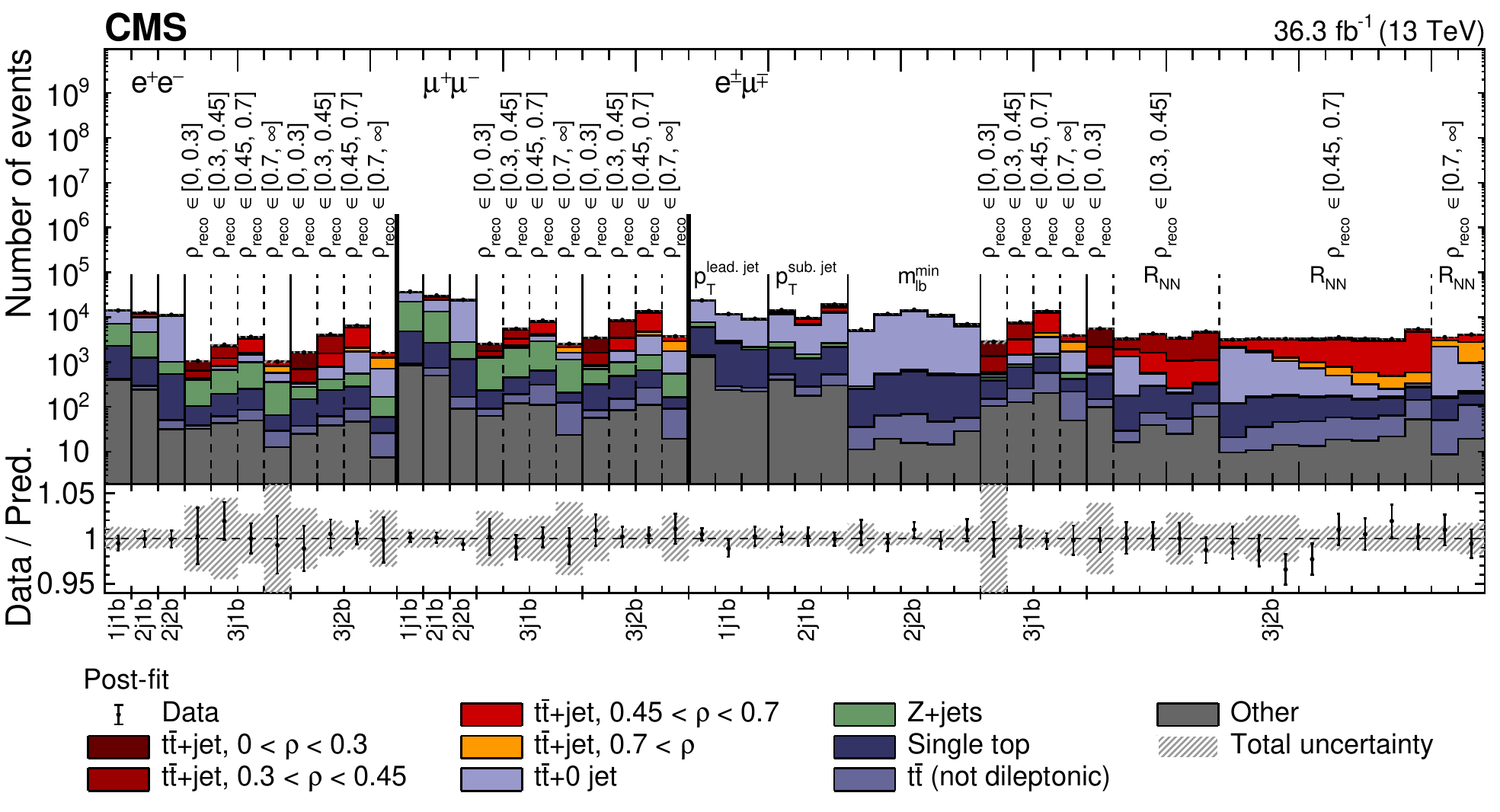}
\caption{The data and MC predicted signal and background yields after the likelihood fit~\cite{CMS:2022emx}.}
\label{fig:postfit}
\end{figure}

Using a profiled likelihood fit, the \ttbarjet\ cross section is directly measured at the parton level. For the unfolding of the $\rho$ distribution, a binning of 0--0.3, 0.3--0.45, 0.45--0.7, and 0.7--1 is considered.
The fitted variables are the relative response of the classifier NN with respect to \ttbarnojet, the jet \pt, the minimal invariant mass of the lepton and b jets, and the total event yield. Events are categorized based on the $\rho$ value, the jet and b jet multiplicities, and the dilepton type. Systematic uncertainties are profiled as nuisance parameters, and the dependence on \mt\ in the MC, \mtmc, is mitigated by adding it as a free parameter. The data and MC predicted signal and background distributions after the fit are shown in Fig.~\ref{fig:postfit}, where a good agreement is observed.
\section{Results and top quark pole mass extraction}
\begin{figure}[htbp]
\centering
\includegraphics[width=0.42\textwidth]{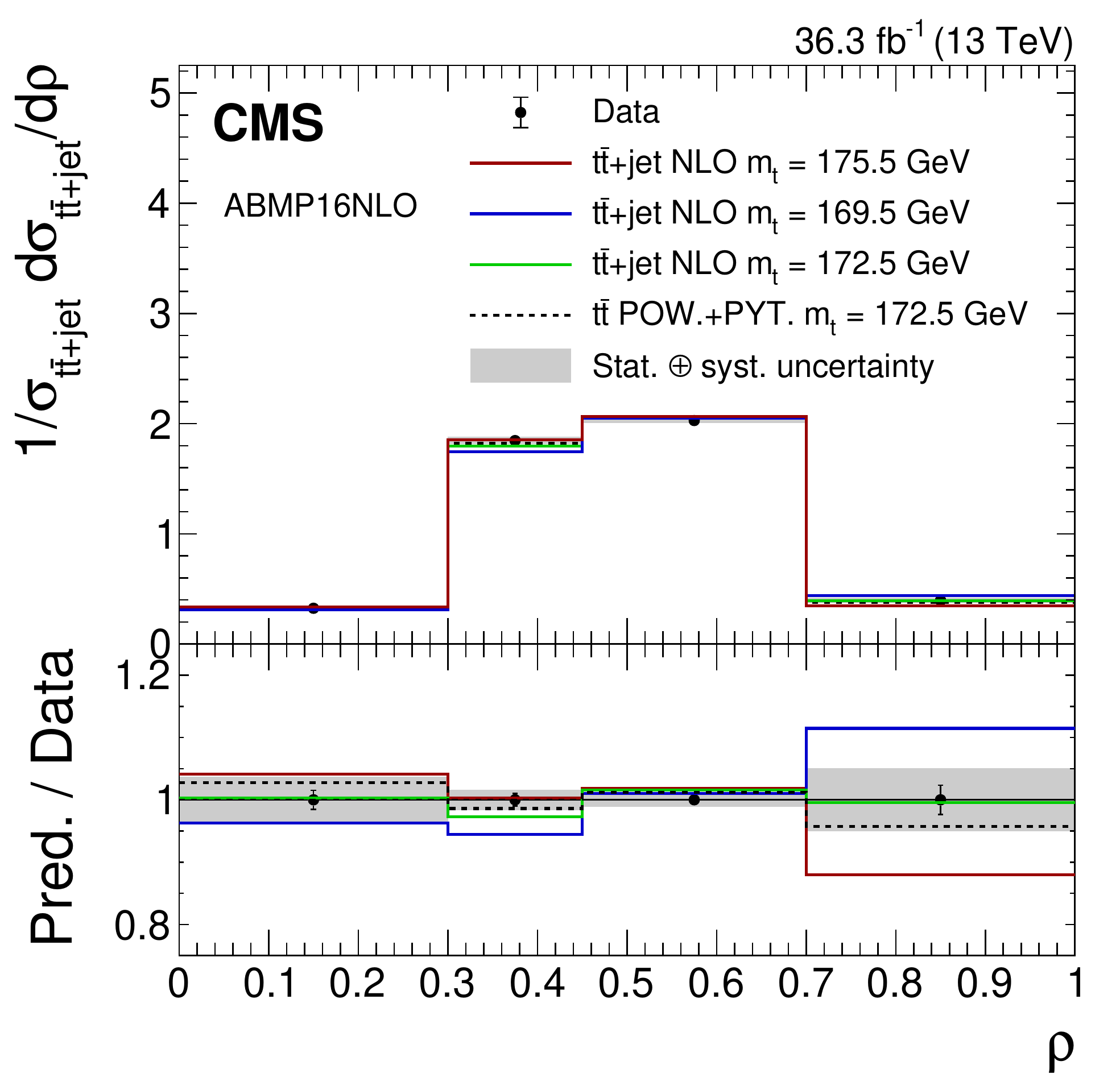}
\includegraphics[width=0.42\textwidth]{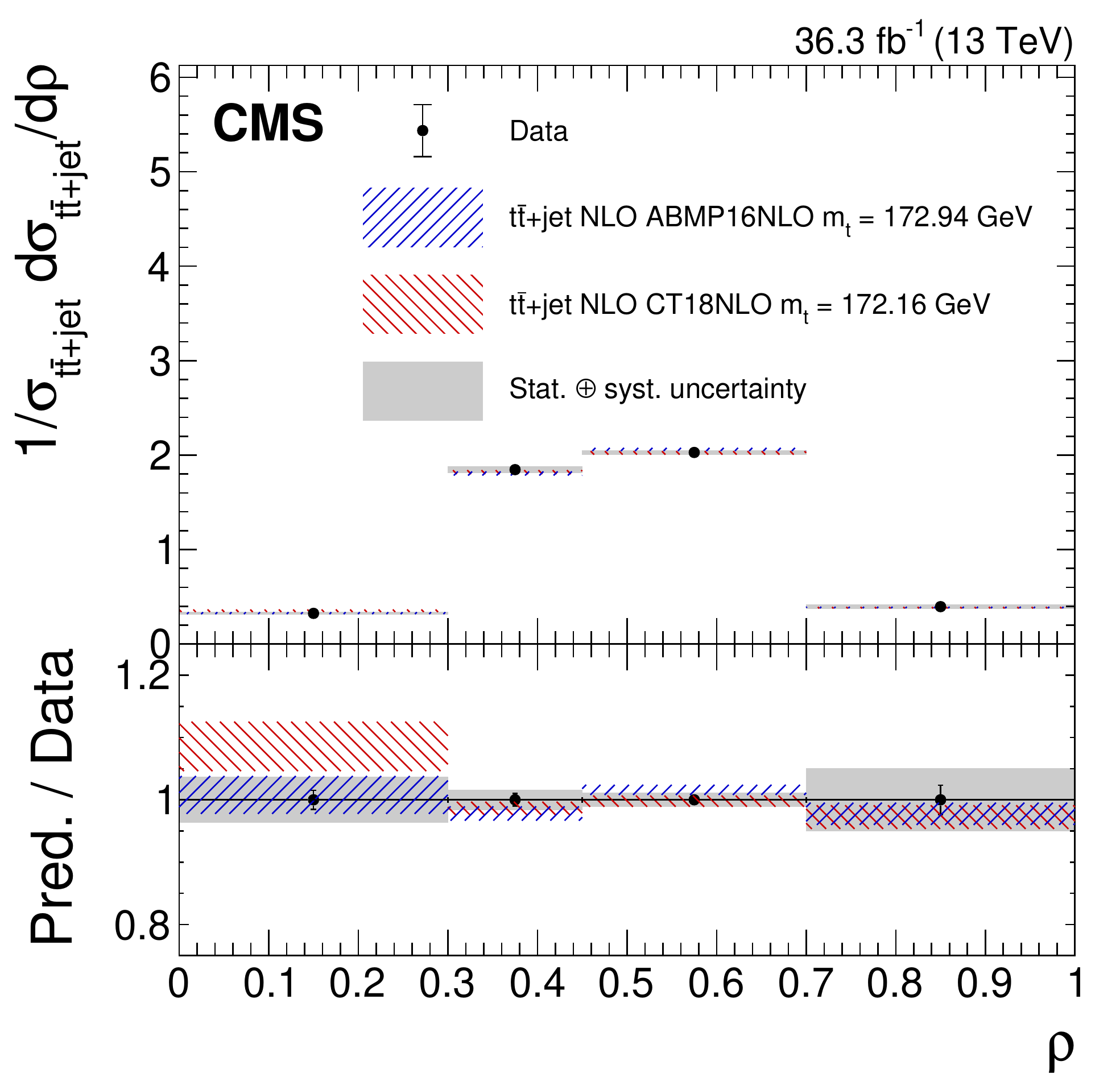}
\caption{Left: The normalized \ttbarjet\ differential cross section as a function of $\rho$ compared to theoretical predictions for values of \mtpole~\cite{CMS:2022emx}. Right: The same measured cross section is shown, but compared to the predictions using the best-fit values of \mtpole~\cite{CMS:2022emx}.}
\label{fig:results}
\end{figure}

The \ttbarjet\ cross section as a function of $\rho$ is directly obtained from the fit, and the normalized distribution is shown in Fig.~\ref{fig:results} left. It is compared to theoretical predictions at NLO for different values of \mtpole. Here, the ABMP16NLO~\cite{bib:ABMP16} parton distribution function (PDF) is used.
By performing a $\chi^2$ fit to the NLO predictions, and taking into account all bin-to-bin correlations in the covariance matrix of the fit, the value of \mtpole\ is extracted. Additionally, PDF uncertainties and uncertainties arising from the extrapolation to the full phase space are included in the fit. The uncertainty due to the choice of the matrix-element scales is externalized.
A value of $\mtpole=172.94\pm1.27\,\text{(fit)}\,^{+0.51}_{-0.43}\,\text{(scale)}$ GeV is measured when using the ABMP16NLO PDF set. For the CT18NLO~\cite{bib:CT18} PDF set, the value is $\mtpole=172.16\pm1.35\,\text{(fit)}\,^{+0.50}_{-0.40}\,\text{(scale)}$ GeV. The measured cross section compared to predictions using the best-fit values for \mtpole\ are shown in Fig.~\ref{fig:results} right. The values of \mtpole\ are in good agreement with previous measurements, also using the same method as done by the ATLAS Collaboration at $\sqrt{s}=8$ TeV~\cite{bib:ttjetAtlas8}.
\bibliographystyle{JHEP}
\bibliography{main}

\providecommand{\href}[2]{#2}\begingroup\raggedright\begin{thebibliography}{1}

\bibitem{bib:hoangmass}
A.H.~Hoang, \emph{What is the top quark mass?},
  \href{https://doi.org/10.1146/annurev-nucl-101918-023530}{\emph{Ann. Rev.
  Nucl. Part. Sci.} {\bfseries 70} (2020) 225}
  [\href{https://arxiv.org/abs/2004.12915}{{\ttfamily 2004.12915}}].

\bibitem{CMS-PAS-TOP-21-008}
{CMS Collaboration}, \emph{{Measurement of the top quark pole mass using
  $\text{t}\overline{\text{t}}\text{+jet}$ events in the dilepton final state
  at $\sqrt{s}=$ 13 TeV}},  CMS Physics Analysis Summary
  \href{https://cds.cern.ch/record/2804936}{CMS-PAS-TOP-21-008}, CERN, Geneva
  (2022).

\bibitem{CMS:2022emx}
{CMS Collaboration}, \emph{{Measurement of the top quark pole mass using
  $\mathrm{t\bar{t}}$+jet events in the dilepton final state in proton-proton
  collisions at $\sqrt{s}$ = 13 TeV}},
  \href{https://arxiv.org/abs/2207.02270}{{\ttfamily 2207.02270}}.

\bibitem{bib:ttjPheno}
S.~Alioli, J.~Fuster, M.V.~Garzelli, A.~Gavardi, A.~Irles, D.~Melini et~al.,
  \emph{Phenomenology of $\ttbar\mathrm{j}+\mathrm{X}$ production at the
  {LHC}}, \href{https://doi.org/10.1007/JHEP05(2022)146}{\emph{JHEP} {\bfseries
  05} (2022) 146} [\href{https://arxiv.org/abs/2202.07975}{{\ttfamily
  2202.07975}}].

\bibitem{bib:ABMP16}
S.~Alekhin, J.~Bl{\"u}mlein and S.~Moch, \emph{{NLO PDFs} from the {ABMP16
  fit}}, \href{https://doi.org/10.1140/epjc/s10052-018-5947-1}{\emph{Eur. Phys.
  J. C} {\bfseries 78} (2018) 477}
  [\href{https://arxiv.org/abs/1803.07537}{{\ttfamily 1803.07537}}].

\bibitem{bib:CT18}
T.-J.~Hou, K.~Xie, J.~Gao, S.~Dulat, M.~Guzzi, T.J.~Hobbs et~al., ``Progress in
  the {CTEQ-TEA NNLO global QCD analysis}.'' 2019.

\bibitem{bib:ttjetAtlas8}
{ATLAS Collaboration}, \emph{Measurement of the top-quark mass in
  $\ttbar+1$-jet events collected with the {ATLAS} detector in pp collisions at
  $\sqrt{s} = {8\text{TeV}}$},
  \href{https://doi.org/10.1007/JHEP11(2019)150}{\emph{JHEP} {\bfseries 11}
  (2019) 150} [\href{https://arxiv.org/abs/1905.02302}{{\ttfamily
  1905.02302}}].

\end{thebibliography}\endgroup
\end{document}